\documentstyle[preprint,aps]{revtex}
\renewcommand{\theequation}{\arabic{section}.\arabic{equation}}
\tighten
\begin{document}
\draft 
\title{The effect of the ultraviolet part of the gluon propagator on 
the heavy quark propagator} 
\author{C.\ J.\ Burden \vspace*{0.2\baselineskip}}
\address{
Department of Theoretical Physics,
Research School of Physical Sciences and Engineering,
Australian National University, Canberra, ACT 0200, Australia
\vspace*{0.2\baselineskip}\\}
\maketitle
\begin{abstract}
We revisit our recently proposed formalism for dealing with the heavy 
quark limit of the quark Dyson-Schwinger equation.  An ambiguity inherent 
in the original version of method is identified and resolved.  Our reanalysis 
illustrates the importance of correctly accounting for the effect of 
hard gluon momenta on the heavy quark self energy in the vicinity of 
the bare fermion mass pole.  
\end{abstract}
%-----------------------------------------------------------------
\setcounter{equation}{0}
\section{Introduction}

In a previous paper\cite{B98} we have developed a formalism for dealing 
with the quark Dyson-Schwinger equation (DSE) in the heavy quark limit
$m_R \rightarrow \infty$ where $m_R$ is the renormalised heavy quark 
mass.  The purpose of this brief report is to point out an ambiguity in 
the limiting procedure described in the previous paper, and to illustrate 
the procedure for extracting the correct heavy quark limit.  The 
ambiguity arises from failing to take proper account of the contribution 
of hard gluon momenta to the loop integral in the DSE.  We begin with 
a brief summary of the renormalised DSE.  

\setcounter{equation}{0}
\section{Dyson-Schwinger equation} 

Our starting point is the renormalised quark DSE~\cite{RW94} 
\begin{equation}
\Sigma'(p,\Lambda) = Z_1(\mu^2,\Lambda^2)\frac{4g^2}{3} 
 \int^\Lambda \frac{d^4q}{(2\pi)^4}\, 
D_{\mu\nu}(p - q) \gamma_\mu S(q) \Gamma_\nu(q,p),   \label{DSEqn}
\end{equation}
where we have used a Euclidean metric in which timelike vectors satisfy 
$p^2 = -p_{\rm Minkowski}^2 < 0$, 
and for which $\{\gamma_\mu,\gamma_\nu\} = 2\delta_{\mu \nu}$.  
We write the renormalised quark propagator in the form
\begin{equation}
S(p,\mu) = \frac{1}{i\gamma\cdot p A(p^2,\mu^2) + B(p^2,\mu^2)} 
         = -i\gamma\cdot p \sigma_V(p^2,\mu^2) + \sigma_S(p^2,\mu^2),  
\end{equation}
where
$\sigma_V(p^2) = A(p^2)/(p^2 A^2 + B^2)$ and 
$\sigma_S(p^2) = B(p^2)/(p^2 A^2 + B^2)$.
The unrenormalised self energy is written
\begin{equation} 
\Sigma'(p,\Lambda) = i\gamma\cdot p \left[A'(p^2,\Lambda^2) - 1\right]
       + B'(p^2,\Lambda^2),  \label{Sigpr}
\end{equation}
where renormalised and bare quantities are related by 
\begin{equation}
 A(p^2,\mu^2) = 1 + A'(p^2,\Lambda^2) - A'(\mu^2,\Lambda^2), \label{AApr}
\end{equation}
\begin{equation}
 B(p^2,\mu^2) = m_R(\mu^2) + B'(p^2,\Lambda^2) - B'(\mu^2,\Lambda^2). 
                     \label{BBpr}
\end{equation}
The renormalisation scale is set such that 
\begin{equation}
\left.S(p)^{-1}\right|_{p^2 = \mu^2}=i\gamma\cdot p + m_R(\mu^2).
                            \label{mufix}
\end{equation}
The set of equations (\ref{DSEqn}) to 
(\ref{mufix}) together with the `abelian approximation' $Z_1 = Z_2$ can 
be solved numerically for the propagator functions $A$ and $B$ once the 
renormalised quark-gluon vertex function $\Gamma_\mu$, renormalised 
gluon propagator $D_{\mu \nu}$, renormalisation point 
$\left(\mu,m_R(\mu)\right)$ and cutoff $\Lambda$ are specified.  

In a general covariant gauge, the gluon propagator takes the form 
\begin{equation}
g^2 D_{\mu \nu}(k) = 
  \left(\delta_{\mu \nu} - \frac{k_\mu k_\nu}{k^2}\right) \Delta(k^2) 
        + g^2 \xi \frac{k_\mu k_\nu}{k^4},    \label{gprop}
\end{equation}
where $\xi$ is the gauge fixing parameter.  In this note we shall assume the 
bare vertex or ``rainbow'' approximation for the vertex function, namely 
$\Gamma_\mu(p,q) = \gamma_\mu$.  

Using Dirac trace identities to project out from Eqs.~(\ref{DSEqn}) 
to (\ref{Sigpr}) a pair of coupled integral equations gives 
\begin{eqnarray}
A'(p^2,\Lambda^2) & = & 
 1 + \frac{4Z_1}{3p^2}  
  \int^\Lambda \frac{d^4q}{(2\pi)^4}\, \left\{ \left[p\cdot q + 
  2\frac{p\cdot(p - q) q\cdot(p - q)}{(p - q)^2}\right]\right. 
      \Delta\left[(p - q)^2\right]                 \nonumber \\
 & &  \left. + \xi  \left[p\cdot q - 
  2\frac{p\cdot(p - q) q\cdot(p - q)}{(p - q)^2}\right]
  \frac{g^2}{(p - q)^2}\right\} \sigma_V(q^2),  \label{ASDE}
\end{eqnarray}
\begin{equation}
B'(p^2,\Lambda^2) = \frac{4}{3}Z_1  \int^\Lambda \frac{d^4q}{(2\pi)^4}\, 
  \left\{3 \Delta\left[(p - q)^2\right] + \xi\frac{g^2}{(p - q)^2}\right\}
      \sigma_S(q^2).  \label{BSDE}
\end{equation}

\setcounter{equation}{0}
\section{The heavy quark limit} 

The heavy quark limit is taken by assuming the expansions 
\begin{equation}
A(p^2,\mu^2) = 1 + \frac{\Sigma_A(K,\kappa)}{m_R(\mu^2)},
\hspace{5 mm}
B(p^2,\mu^2) = m_R(\mu^2) + \Sigma_B(K,\kappa),  \label{sigABdef}
\end{equation}
where we have defined the new momentum variable $K$ and renormalisaton 
point $\kappa$ by 
\begin{equation}
K = \frac{p^2 + m_R^2}{2im_R}, \hspace{5 mm}
\kappa = \frac{\mu^2 + m_R^2}{2im_R}.   \label{Kdef}
\end{equation}
The change of independent variable $p^2 \rightarrow K$ 
is illustrated in Fig.~1 of Ref.\cite{B98}.  Our aim is to solve for 
the quark propagator in the vicinity of $p^2 = -m_R^2$, or equivalently, 
$K = 0$.  From Eq.~\ref{sigABdef} we have 
\begin{equation}
m_R\sigma_V(p^2) = \frac{1}{2} \sigma_Q(K) + O\left(\frac{1}{m_R}\right), 
                    \hspace{5 mm}
\sigma_S(p^2) = \frac{1}{2} \sigma_Q(K) + O\left(\frac{1}{m_R}\right), 
                        \label{sigexp}
\end{equation}
where we have defined 
\begin{equation}
\sigma_Q(K) = \frac{1}{i K + \Sigma(K)}, 
                    \hspace{5 mm}\mbox{ and } \hspace{5 mm}
\Sigma(K) = \Sigma_B(K) - \Sigma_A(K).  \label{Sigdef}
\end{equation}  
Note that Eq.~(\ref{sigexp}) is valid in any part of the complex momentum 
plane in which Eq.~(\ref{sigABdef}) holds, that is, 
there is no assumption that we are in the vicinity of $K=0$.  

For the purposes of expanding the DSE in the inverse quark mass we 
introduce the momentum substitutions generic to heavy quark effective 
theory 
\begin{equation}
p_\mu = imv_\mu + k_\mu, \hspace{5 mm} q_\mu = imv_\mu + k'_\mu, 
                         \label{kmudef}
\end{equation}
where $v_\mu v_\mu = 1$.  Typically we take $v_\mu = (0,0,0,1)$, which 
entails $K = k_4 + O(1/m_R)$ in the vicinity of $p^2 = -m_R^2$.  
Note also that the result $S(p,\mu) = \frac{1}{2}(1 + \gamma_4) 
\sigma_Q(K,\kappa) + O(1/m)$ for the dressed quark propagator 
is only a good approximation for $\left|k_\mu\right| << m_R$.  

In Ref.~\cite{B98} the heavy quark DSE is obtained by substituting 
Eqs.(\ref{sigexp}) and (\ref{kmudef}) into Eqs.(\ref{ASDE}) and 
(\ref{BSDE}), and then using Eqs.~(\ref{AApr}) and (\ref{BBpr})
to obtain an integral equation 
for the heavy quark self energy $\Sigma(K)$.  The procedure involves 
truncating a $1/m_R$ expansion of the integrand of the DSE.  However, 
for heavy quarks, 
and for a model gluon propagator with a realistic ultraviolet tail, 
the DSE receives contributions from values of $k'_\mu$ (defined in 
Eq.~(\ref{kmudef})) which are not small compared with $m_R$.  Because of 
this the $1/m_R$ expansion must be applied with some caution.  

To understand in detail what can go wrong with the procedure, 
consider the model gluon propagator 
used in ref.\cite{B98}, namely the smeared Frank and Roberts propagator 
\begin{equation}
\Delta(k^2) = (2\pi)^4 \frac{m_t^2 d}{\alpha^2 \pi^2}
      e^{-k^2/\alpha} + 4\pi^2 d \frac{1 - e^{-k^2/(4m_t^2)}}{k^2}
      = \Delta_{\rm IR}(k^2) + \frac{4 \pi^2 d}{k^2}, 
                                          \label{GFRprop}
\end{equation}
where $d = 12/(33 - 2N_f)$, $N_f = 3$ is the number of light quark 
flavours, $m_t = 0.69$ GeV is a parameter fitted to a range 
of calculated pion observables, and the Gaussian width $\alpha$ was chosen 
to be 0.5643 (GeV)$^2$.  As we have indicated, this propagator 
has the appearance of a Coulomb propagator beyond the scale of the 
parameter $m_t$.  Eq.~(\ref{ASDE}) then becomes 
\begin{eqnarray}
A'(p^2,\Lambda^2) & = & 
 1 + \frac{4Z_1}{3p^2}  
  \int^\Lambda \frac{d^4q}{(2\pi)^4}\, \left\{ \left[p\cdot q + 
  2\frac{p\cdot(p - q) q\cdot(p - q)}{(p - q)^2}\right] \right. 
      \Delta_{\rm IR}\left[(p - q)^2\right] \sigma_V(q^2) \nonumber \\
& & \left .+ 4\pi^2 d \left[(1 + \xi')p\cdot q +  
  2(1 - \xi')\frac{p\cdot(p - q) q\cdot(p - q)}{(p - q)^2}\right]  
           \frac{ \sigma_V(q^2)}{(p - q)^2}\right\},   \label{ASDE2}
\end{eqnarray}
where $\xi' = g^2\xi/(4\pi^2 d)$.
The procedure employed in Ref.\cite{B98} of taking the heavy quark limit 
by substituting Eq.~(\ref{kmudef}) and keeping only leading order in 
$1/m_R$ contributions to the integrand is certainly 
valid for the term containing 
$\Delta_{\rm IR}$.  This is because the exponentially damped gluon 
propagator ensures that the factor in square brackets is only sampled for 
$k'_\mu << m_R$.  For the remaining Coulomb part, however, one runs into 
problems.  

Consider for a moment the identity
\begin{equation}
0 = \int \frac{d^4q}{(2\pi)^4}\,  \left[p\cdot q +  
  2\frac{p\cdot(p - q) q\cdot(p - q)}{(p - q)^2}\right] 
                   \frac{ \sigma_V(q^2)}{(p - q)^2} .  \label{iden}
\end{equation}
One easily checks that the integrand of Eq.~(\ref{iden}) is equal to 
\begin{equation}
I(\left|p\right|,\left|q\right|,\theta) = 
\frac{\left|p\right|\left|q\right|\sigma_V(q^2)}{\sin^2\theta}
  \frac{d}{d\theta}\left[\frac{\sin^3\theta}{(p - q)^2}\right], 
\end{equation}
where $\left|p\right| = (p\cdot p)^{1/2}$, 
$\left|q\right| = (q\cdot q)^{1/2}$ and $\theta$ is the angle between 
$p_\mu$ and $q_\mu$ in Euclidean 4-space.  Clearly the angular part of 
the integral, namely $\int I(\left|p\right|,\left|q\right|,\theta) 
\sin^2\theta d\theta$, is identically zero.  In Landau gauge, $\xi' = 0$, 
the second term in Eq.~(\ref{ASDE2}) should therefore integrate to 
zero because of cancellations of significant contributions from all 
directions in Euclidean momentum space.  In terms of the Feynman diagram 
corresponding to the original Euclidean DSE, Eq.~(\ref{DSEqn}), this means 
that in the heavy quark limit hard momentum from the incoming quark line 
can flow through the gluon propagator line, significantly altering the 
direction of the internal quark line.  Considerable cancellations occur
in the integration over angles, resulting in a diminished net contribution 
to Eq.~(\ref{ASDE2}).  

Inserting the change of momentum variables Eq.~(\ref{kmudef}) into the last 
term in Eq.~(\ref{ASDE2}) we shall see below that these cancellations do not 
necessarily occur order by order in $1/m_R$.  In order to construct a 
viable formalism for extracting the heavy quark limit, it is therefore 
necessary to subtract off the angular oscillatory behaviour of the 
integrand before expanding in the inverse quark mass.  

To this end, we add to Eq.~(\ref{ASDE2}) an arbitrary multiple of the 
identity Eq.~(\ref{iden}) to give 
\begin{equation}
A'(p^2,\Lambda^2) =  1 + \mbox{(infrared part)} 
      + \frac{16\pi^2 Z_1 d}{3p^2}  
  \int^\Lambda \frac{d^4q}{(2\pi)^4}\, H(p,q)  
                   \frac{ \sigma_V(q^2)}{(p - q)^4} , \label{ASDEsub}
\end{equation}
where 
\begin{equation}
 H(p,q) = (1 + \xi' + \lambda)p\cdot q (p - q)^2 
    + 2(1 - \xi' + \lambda)p\cdot(p - q) q\cdot(p - q),  \label{Hdef}
\end{equation}
with $\lambda$ arbitrary.  

We can now make use of Eqs.~(\ref{sigexp}) and (\ref{kmudef}) to extract 
from Eqs.~(\ref{AApr}), (\ref{BBpr}), (\ref{BSDE}) and (\ref{ASDEsub}) 
an integral equation for $\Sigma(K,\kappa)$.  
For example, in Landau gauge ($\xi = 0$), and choosing $k_\mu = (0,0,0,K)$ 
(which is allowable to zeroth order in $1/m_R$ provided $K<<m_R$), we obtain
\begin{eqnarray}
\Sigma(K,\kappa) & = & \frac{4}{3} \int^\Lambda \frac{d^4k'}{(2\pi)^4}\, 
     \frac{1}{ik_4' + k'^2/(2m_R) + \Sigma(k'_4,\kappa)} \nonumber \\
 & & \times \left( \left\{ \frac{\left|{\bf k}'\right|^2 
       \Delta_{\rm IR}\left[(K - k'_4)^2 + \left|{\bf k}'\right|^2 \right]}
 {(K - k'_4)^2 + \left|{\bf k}'\right|^2} \right. \right. -
                                                         \nonumber \\
 & & \left. \left. \left[\frac{3\lambda}{2} - (1 + \lambda) 
  \frac{\left|{\bf k}'\right|^2}{(K - k'_4)^2 + \left|{\bf k}'\right|^2}
      \right] \frac{4\pi^2 d}{(K - k'_4)^2 + \left|{\bf k}'\right|^2}
               \right\} - (K \rightarrow\kappa) \right).    \label{hsde2}
\end{eqnarray}  
The $\lambda = 0$ case was solved numerically in Ref.\cite{B98}.  
For arbitrary $\lambda$, however, we find that the solutions are heavily 
dependent on the irrelevant parameter $\lambda$, so something is clearly 
amiss.  

As pointed out above, and in the Appendix to Ref.\cite{B98}, the integrand 
of this equation is sampled for values of $k'$ which are not necessarily 
small compared with $m_R$.  The terms $\sigma_S(q^2)/(p - q)^2$ and 
$\sigma_V(q^2)/(p - q)^4$ in the Coulomb parts of integrands 
of Eqs.~(\ref{BSDE}) and (\ref{ASDEsub}) cannot be the cause of the 
problem, since, 
as pointed out below Eq.~(\ref{sigexp}), they are well represented by 
the heavy quark approximations $\frac{1}{2}\sigma_Q/(k - k')^2$ and 
$\frac{1}{2}\sigma_Q/[m_R(k - k')^4]$ throughout the complex momentum plane.  
The problem must therefore lie solely with truncation of the $1/m_R$ 
expansion of $H(p,q)$ in the integrand of Eq.~(\ref{ASDEsub}).  By 
careful choice of the parameter $\lambda$ however, we find that we can 
subtract of the troublesome 
angular oscillatory behaviour hidden in the identity (\ref{iden}) to 
leave an integrand which can be accurately expanded in the inverse quark 
mass even for large values of $k'$.  

In the Appendix we show that, with $k_\mu = (0,0,0,K)$, the parameter 
choice 
\begin{equation}
\lambda = \xi/3 - 1         \label{cor}
\end{equation} 
implies 
\begin{equation}
H(p,q) = -\frac{4\xi'm_R^2}{3} \left|{\bf k}'\right|^2
              \left[1+O\left(\frac{K}{m_R}\right)\right].  
\end{equation}
We are interested in solving the DSE equation in the vicinity of 
the bare quark mass pole, that is, for $\left|K\right| << m_R$.  In 
this case, as $m_R\rightarrow \infty$, the non-leading terms in the 
$1/m_R$ expansion of $H$ do not contribute to the integral in 
Eq.~(\ref{ASDEsub}) and our procedure for obtaining the heavy quark 
limit is valid.  For any other value of $\lambda$ however, the 
$1/m_R$ expansion of $H$ picks up $O(k'/m_R)$ pieces which contribute 
to the ultraviolet part of the integral and cannot be neglected.  This 
is the source of the discrepancy.  

\begin{figure}[t]
\caption{The modulus $\left|\sigma_Q(K)\right|$ of the heavy quark 
obtained by solving the heavy quark 
DSE Eq.~(\ref{hsde2}) in rainbow approximation with a Landau gauge, 
gaussian smeared Frank and Roberts gluon propagator.  
The renormalised heavy quark 
mass is $m_R = 5$ GeV, and remaining input parameter values are given 
in the text.  The steeper surface with an obvious pole is the solution 
found in Ref.\protect\cite{B98} corresponding to $\lambda = 0$.  The 
shallower surface is the solution corresponding to 
the correct parameter value, $\lambda = - 1$.
\label{fig1} } 
\end{figure}
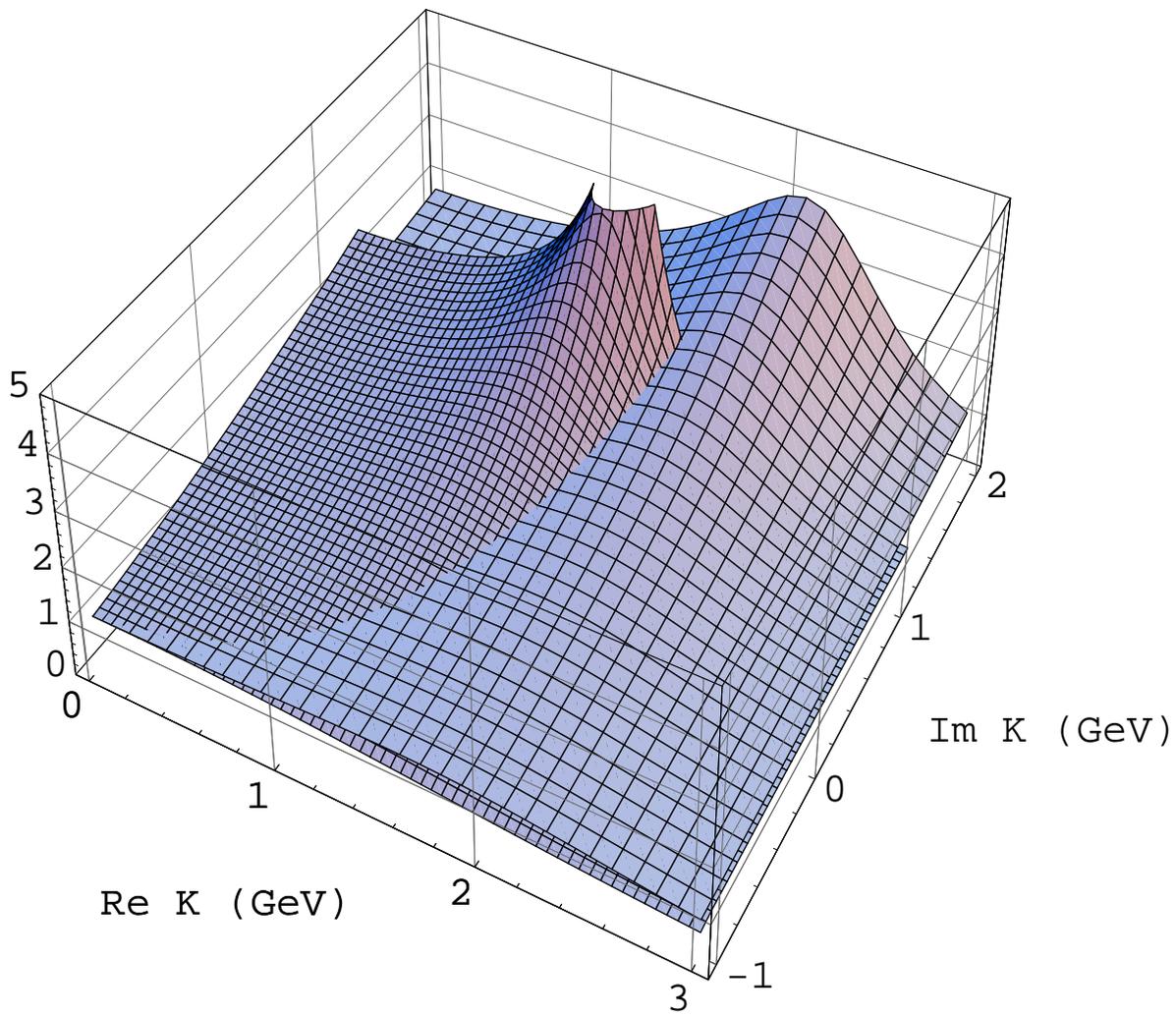

In Fig.~\ref{fig1} is plotted the modulus of the heavy quark propagator 
obtained by solving Eq.~(\ref{hsde2}) in Landau gauge with parameter 
value $\lambda = -1$, consistent with Eq.~(\ref{cor}). 
For comparison, we also plot the solution from Ref.\cite{B98} corresponding 
to $\lambda = 0$.  With the correct value of $\lambda$, conjugate poles 
have moved considerably deeper into the complex momentum plane, and 
away from the imaginary $K$ axis, or equivalently, away from the real, 
timelike $p^2$ axis.  Very little evidence is left of the original 
bare propagator mass pole at $K=0$.  

\setcounter{equation}{0}
\section{Comment on Euclidean and Minkowski metrics}

Our formalism presupposes the quark DSE to be formulated in Euclidean 
space.  This is standard procedure in most nonperturbative treatments of 
confining field theory including DSE calculations~\cite{RW94}.  
Opinions differ regarding the question of whether this procedure is 
simply a matter of practical expediency, or whether it reflects a deeper 
truth about the nature of confining theories.  In its most extreme 
form~\cite{SC90}, the position adopted is that the underlying 
quark degrees of freedom can only be meaningfully considered in Euclidean 
space.  
If we adopt a less extreme point of view, it may be instructive 
to explore whether the angular oscillatory function in the 
integrand of Eq.~(\ref{iden}) finds its analogue in a Minkowski formalism 
of the quark DSE.  That is, does the quark self energy contain significant 
contributions from different directions in physical Minkowski space time 
which cancel?  

Consider the case where the Minkowski space vectors $p^\mu$ and $q^\mu$ 
are both timelike.  We define $\left|p\right| = (p^\mu p_\mu)^{1/2}$ and 
$\left|q\right| = (q^\mu q_\mu)^{1/2}$.  Choosing 
$p^\mu = (\left|p\right|,{\bf 0})$ and 
$q^\mu = (\eta\left|q\right|\cosh\theta,
\hat{\bf q} \left|q\right|\sinh\theta)$ 
where $\eta = \pm 1$ and $\theta$ is real, the integrand of 
Eq.~(\ref{iden}) becomes 
%\begin{eqnarray}
%\lefteqn{\left[p\cdot q +  
%  2\frac{p\cdot(p - q) q\cdot(p - q)}{(p - q)^2}\right] 
%                   \frac{ \sigma_V(q^2)}{(p - q)^2} } \nonumber \\
% & = & \frac{\eta  \left| p\right| \left| q\right| \sigma_V(q^2)}
%     {\sinh^2\theta} \frac{d}{d\theta} 
%   \left(\frac{\sinh^3\theta}{\left| p\right|^2 + \left| q\right|^2 
%          - 2 \eta \left| p\right| \left| q\right| \cosh\theta} \right),  
%\end{eqnarray}
\begin{equation}
\left[p\cdot q +  
  2\frac{p\cdot(p - q) q\cdot(p - q)}{(p - q)^2}\right] 
                   \frac{ \sigma_V(q^2)}{(p - q)^2} 
  =  \frac{\eta  \left| p\right| \left| q\right| \sigma_V(q^2)}
     {\sinh^2\theta} \frac{d}{d\theta} 
   \left(\frac{\sinh^3\theta}{\left| p\right|^2 + \left| q\right|^2 
          - 2 \eta \left| p\right| \left| q\right| \cosh\theta} \right),  
\end{equation}
and the integration measure within the forward and backward 
light cones becomes 
\begin{equation}
\int d^4 q = \sum_{\eta = \pm 1} 
  \int_0^\infty \left| q\right|^3 d\left| q\right| 
  \int_0^\infty \sinh^2\theta d\theta \int d\Omega. 
\end{equation}
The analogue of the integrand over angles at fixed $\left| q\right|$, which 
is zero in the Euclidean formalism, is now 
\begin{equation}
\sum_{\eta = \pm 1} \int_0^\infty  d\theta 
\frac{d}{d\theta} 
   \left(\frac{\eta\sinh^3\theta}{\left| p\right|^2 + \left| q\right|^2 
          - 2 \eta \left| p\right| \left| q\right| \cosh\theta} \right).  
\end{equation}
It is difficult to conceive of a regularisation procedure which could render 
such a severely divergent integral equal to zero.  A similar analysis of 
the case where $p_\mu$ is timelike and 
$q_\mu$ is spacelike also leads to a strongly divergent 
integral.  We are led to the conclusion that the subtraction procedure 
described in the previous section is only relevant to the Euclidean 
formalism.  

\setcounter{equation}{0}
\section{Conclusions}

We have carried out a reanalysis of the quark DSE in the heavy 
mass limit $m_R \rightarrow \infty$.  
It is observed that, for a model gluon propagator with a realistic 
asymptotic ultraviolet tail, the procedure used in our previous 
analysis\cite{B98} may not lead to a unique solution for the dressed heavy 
quark propagator in the vicinity of the bare fermion mass pole.  The 
problem lies with failing to account adequately for the contribution to 
the DSE loop integral from the ultraviolet part of the gluon propagator.  
In essence, this means that in the Euclidean formalism, the momentum of 
an intermediate quark state cannot be assumed to be aligned closely with 
that of the external quark line in the heavy quark limit.  

We have solved this problem by adding to the integrand of the DSE 
a function which integrates to zero over angles in Euclidean space, 
but which cancels off higher order in $1/m_R$ contributions to the 
integrand that would otherwise contribute to the zeroth order quark self 
energy.  We note that the analogous function in Minkowski space, when 
integrated over the two sheeted hyperboloid defined by a constant value for 
$q^\mu q_\mu$, is a strongly divergent integral, suggesting that our 
procedure is unique to the Euclidean formalism.  In fact, 
if such a procedure proved to be necessary  
and valid in Minkowski space, it would bring into question the 
notion that backward propagating intermediate states can be ignored  
in the heavy quark limit.  In the absence 
of a viable nonperturbative treatment of quantum chromodynamics directly 
formulated in Minkowski space, it is difficult to place a physical 
interpretation on a procedure which appears to be simply a necessary 
mathematical step within the Euclidean formalism. 

The practical effect of the correction is to shift conjugate propagator poles 
present in our previous solution deeper into the complex momentum plane, and 
further from the real, timelike axis.  Our solution for the heavy quark 
propagator can be regarded as the first step in determining the 
Bethe-Salpeter (BS) amplitude for $Q$-$\bar{q}$ mesons.  In reference 
\cite{BL97} it was found that, if a simple Gaussian gluon propagator which 
does not accurately 
model the ultraviolet gluon behaviour is used, the pole in the heavy quark 
propagator is only slightly shifted from its bare propagator position 
at $K = 0$.  It was further shown that the presence of this pole prevented 
solutions to the ladder approximation $Q$-$\bar{q}$ BS equation.  
The removal of propagator poles to remote parts of the complex plane 
exhibited by our corrected solutions should overcome this difficulty.  

A calculation of the Isgur-Wise function using an impulse approximation 
involving DSE inspired light quark propagators and bare heavy quark 
propagators has been carried out in reference \cite{IKMR98}.  In this 
calculation the bound state $Q$-$\bar{q}$ BS amplitude was modelled 
using a variety of Gaussian-like functional forms, each using a single 
fitted scale parameter.  It was found that the calculated Isgur-Wise 
function is insensitive to which functional form used provided a suitable 
choice was made for the scale parameter.  The appropriateness of the fitting 
procedure employed in reference \cite{IKMR98} 
can be tested once the next step of modelling the BS equation 
with the corrected heavy quark propagator is completed.

\section*{Acknowledgement}

The author acknowledges helpful discussions with F.\ T.\ Hawes and 
C.\ D.\ Roberts, and the hospitality of the Special Research Centre 
for the Subatomic Structure of Matter in Adelaide.  

%---------------------------------------------------------------------------

\renewcommand{\theequation}{\Alph{section}.\arabic{equation}}
\setcounter{section}{1}
\setcounter{equation}{0}
\section*{Appendix: Expansion of $H$}

Inserting Eq.~(\ref{kmudef}) into Eq.~(\ref{Hdef}) we obtain
\begin{equation}
H(p,q) = H_2(p,q) m_R^2 + H_1(p,q) m_R + H_0(p,q), 
\end{equation}
where, after setting $k_\mu = (0,0,0,K)$, 
\begin{eqnarray*} 
H_2 & = & (1 + \lambda)\left[-3(k'_4 - K)^2 - \left|{\bf k}'\right|^2\right] 
      + \xi'\left[(k'_4 - K)^2 - \left|{\bf k}'\right|^2\right] ,  \\
H_1 & = & i(1 + \lambda)
           \left[3K^3 - 3K^2k'_4 -  Kk'^2 + 3k'^2k'_4 - 2K{k'_4}^2\right] \\
& & +i\xi' \left[-K^3 +  K^2k'_4 + 3Kk'^2 -  k'^2k'_4 - 2K{k'_4}^2\right], \\
H_0 & = & (1 + \lambda)
         \left[3K^3k'_4 + 3Kk'^2k'_4 - 2K^2k'^2 - 4K^2{k'_4}^2\right]
 + \xi'  \left[-K^3k'_4 -  Kk'^2k'_4 + 2K^2k'^2\right].  
\end{eqnarray*}
The $k'^2k'_4$ term in $H_1$ and the $Kk'^2k'_4$ in $H_0$ can be disposed 
of by choosing $\lambda = \xi/3 - 1$, which reduces $H(p,q)$ to 
\begin{equation} 
H(p,q) = - \frac{4\xi'm_R^2}{3} \left|{\bf k}'\right|^2
   \left(1 - \frac{2iK}{m_R} - \frac{K^2}{m_R^2}\right).
\end{equation}
For the region $\left|K\right| << m_R$, the final two terms can be ignored 
without affecting the solution of the DSE to zeroth order in $1/m_R$.  
%
%--------------------------------------------------------------------------
%   REFERENCES

%
\pagebreak
\end{document}